# Bridging Stratification and Regression Adjustment: Batch-Adaptive Stratification with Post-Design Adjustment in Randomized Experiments

Zikai Li*

October 26, 2025

**Abstract**

To increase statistical efficiency in a randomized experiment, researchers often use stratification (i.e., blocking) in the design stage. However, conventional practices of stratification fail to exploit valuable information about the predictive relationship between covariates and potential outcomes. In this paper, I introduce an adaptive stratification procedure for increasing statistical efficiency when some information is available about the relationship between covariates and potential outcomes. I show that, in a paired design, researchers can rematch observations across different batches. For inference, I propose a stratified estimator that allows for nonparametric covariate adjustment. I then discuss the conditions under which researchers should expect gains in efficiency from stratification. I show that stratification complements rather than substitutes for regression adjustment, insuring against adjustment error even when researchers plan to use covariate adjustment. To evaluate the performance of the method relative to common alternatives, I conduct simulations using both synthetic data and more realistic data derived from a political science experiment. Results demonstrate that the gains in precision and efficiency can be substantial.

*PhD Candidate, Department of Political Science, University of Chicago. I'm grateful to Alexander Coppock, Andrew Eggers, Robert Gulotty, Molly Offer-Westort and Anton Strezhnev for their feedback on earlier drafts of the paper. I thank the Social Sciences Computing Services at the University of Chicago for computational support. For helpful comments, I thank participants at the 2025 American Causal Inference Conference, the 2024 American Political Science Association Annual Meeting, the 2024 Midwest Political Science Association Annual Meeting, the 2024 Meeting of the Society for Political Methodology, and workshops at the University of Chicago. All errors are my own.

# 1  Introduction

Randomized experiments have become increasingly popular for estimating causal effects in the social sciences. They allow researchers to infer treatment effects with minimal assumptions: under random assignment, the simple difference in means between treated and control groups in the study sample provides an unbiased estimate of the average population-level treatment effect. However, in social science experiments, the precise estimation of treatment effects presents a ubiquitous challenge given limited sample sizes and a lack of understanding of what predicts the outcome prior the experiment. While randomization ensures unbiasedness, it does not guarantee precision or efficiency, particularly when sample sizes are modest. An important goal of experimental design is thus to maximize information gained from each observation. Broadly, there are two ways to improve statistical efficiency: stratification (or blocking) in the design stage and covariate adjustment in the analysis stage. In conventional practice, only covariate adjustment uses information about the predictive relationship between covariates and potential outcomes.

This paper introduces a new method, batch-adaptive pair matching (BAPM) with regression adjustment (BAPM+), that bridges stratification and regression adjustment by integrating predictive modeling into the design stage and using regression adjustment in the analysis stage. The method divides the experiment into batches and uses machine learning models trained on data from earlier batches to guide stratification decisions for subsequent batches. Researchers are often wary of using data on realized outcomes to inform treatment assignment in later batches, a practice that complicates inference in outcome-adaptive designs (Offer-Westort et al. 2021). The treatment probability in BAPM+, however, is fixed ahead of time, and treatment assignment stays independent of potential outcomes. With crossfitting, even though the assignment the algorithm makes for each unit is informed by the data from other units, it does not depend on the realized or potential outcomes for the pair under consideration. The key innovation is thus that researchers can use outcome data from initial batches to inform treatment assignment without compromising inferential validity.

Stratification divides the sample into groups prior to randomization and uses complete randomization within each group for treatment assignment. This approach reduces variance when units within each stratum have similar potential outcomes. However, when dealing with covariates for stratification, researchers face several challenges that can compromise the optimality of their design. When some covariates are continuous, researchers must decide how to partition the covariate space into meaningful strata, often with little principled guidance on cutpoints. With multiple discrete covariates, the number of possible strata grows rapidly, often forcing researchers to select only a subset of variables for stratification. A more flexible approach uses distance measures, such as the Mahalanobis distance, to match units that are "close" in covariate space ([Moore 2012](#)), which accommodates continuous variables and multiple covariates. Put simply, Mahalanobis distance offers a way to measure how close two units are in the covariate space. It does this by rescaling the covariates according to their variances and covariances. These workarounds often fail to fully utilize available covariate information. In many cases we have or can obtain information about the predictive relationship between covariates and outcomes. In such situations, any method that stratifies based only on information about the joint distribution of covariates (like Mahalanobis distance) misses out on potentially substantial efficiency gains.

Consider a simple illustration with two binary covariates: `urban_residence` (coded as 1 if the respondent lives in an urban area and 0 otherwise) and `college` (coded as 1 for having a college degree and 0 otherwise). Suppose we have past data suggesting that urban residence is highly predictive of potential outcomes while college education is largely irrelevant for the outcome of interest. Using Mahalanobis distance matching, a college-educated urban resident (1, 1) and a non-urban resident without a college degree (0, 0) would be considered equally good matches for a non-college-educated urban resident (1, 0), since both lie the same Mahalanobis distance from (1, 0). However, if urban residence strongly predicts outcomes but college education does not, the non-urban resident is likely to have very different potential outcomes than the urban resident. Meanwhile, the college-educated urban resident may resemble the non-college-educated urban resident in expected outcomes, making it the better match. Matching only on covariates cannot exploit this difference in predictive power, leading to suboptimal pairings when outcome-relevant information is available.

Researchers may ask whether stratification is beneficial at all given the availability of regression adjustment, which does use information about outcome-covariate relationships: if regression adjustment can improve precision, why not simply use complete randomization and address any covariate imbalance through regression adjustment in the analysis stage? While regression adjustment can significantly improve precision over unadjusted estimators when covariates are predictive of potential outcomes (Lin 2013, Wager et al. 2016), it cannot fully compensate for covariate imbalance induced by design choices except in unrealistic scenarios. Substantial imbalance in outcome-relevant covariates introduces additional variance that post-design adjustment cannot completely eliminate unless the fitted regression model is almost as good as the oracle one. Thus, stratification complements rather than substitutes for regression adjustment by insuring against imperfect modelling choices and irreducible learning error.

BAPM with regression adjustment seeks to exploit this complementarity. It continually learns the covariate-outcome relationship and uses this information in both the design and analysis stages. In a two-batch implementation, it begins with outcome-agnostic stratification for the first batch. After collecting outcomes for this batch, the method fits flexible models to predict potential outcomes based on covariates. These models then generate predicted potential outcomes for all units across both batches. Next, BAPM pairs units across batches based on these predictions and assigns treatment status for second-batch units. Finally, researchers can use regression adjustment to gain further precision and efficiency in the analysis stage.

This approach offers at least two advantages. First, unlike conventional stratification,

BAPM leverages the predictive relationship between covariates and outcomes when it is available in both the design and analysis stages. Second, it also improves upon within-batch matching approaches that use this predictive relationship. In contrast to stratification methods that only allow within-batch matching, BAPM can look for optimal pairs regardless of batch membership and further improve matching quality and, by extension, statistical efficiency.

This paper is related to but distinct from existing methods for sequential treatment assignment and data collection in political science. These methods all aim to improve upon complete random assignment but differ in objectives and implementations. Offer-Westort et al. (2021) study outcome-adaptive randomization that dynamically adjusts treatment assignment probabilities to identify the best-performing treatment, using outcome data for discovering the most effective treatment and estimating its effect. They thus have a different goal than BAPM+, which focuses on improving efficiency. Blackwell et al. (2023) and Moore & Moore (2013) both focus on statistical efficiency but use different information: Blackwell et al. (2023) use pilot outcome data to optimize sample allocation across arms but ignores covariates, while Moore & Moore (2013) use sequential covariate matching but their method is agnostic to covariate-outcome relationships. BAPM+ differs from both approaches by using both covariate and outcome information to improve efficiency through adaptive pairing across batches.

Blackwell et al. (2023) use pilot outcome data to optimize sample allocation across arms but ignores covariates, while Moore & Moore (2013) use sequential covariate matching but their method is agnostic to covariate-outcome relationships. BAPM+ differs from both approaches by using both covariate and outcome information to improve efficiency through adaptive pairing across batches.

This paper is related to but distinct from existing methods for sequential treatment assignment and data collection in political science. These methods all aim to improve upon complete random assignment but differ in objectives and implementations. Offer-Westort et al. (2021) study outcome-adaptive randomization that dynamically adjusts treatment assignment probabilities to identify the best-performing treatment, using outcome data for discovering the most effective treatment and estimating its effect. They thus have a different goal than BAPM+, which focuses on improving efficiency. Blackwell et al. (2023) and Moore & Moore (2013) both focus on statistical efficiency but use different information: Blackwell et al. (2023) use pilot outcome data to optimize sample allocation across arms but ignores covariates, while Moore & Moore (2013) use sequential covariate matching but their method is agnostic to covariate-outcome relationships. BAPM+ differs from both approaches by using both covariate and outcome information to improve efficiency through adaptive pairing across batches. The approach builds on the advice in Gerber & Green (2012) to "block on … an index that summarizes the covariates into a single prognostic score" (p. 110), though they did not elaborate on how to construct such an index in practice. BAPM+ operationalizes this idea by learning the prognostic score from data rather than relying on researcher intuition about which covariates to prioritize.

The broader experimental design literature has developed several approaches that share key insights with BAPM+. Bai (2022) demonstrates that pairing units based on the expected sum of their potential outcomes achieves the efficiency bound among all stratified randomization schemes. Athough Bai (2022) focuses on single-stage and within-batch stratification, the theoretical result motivates BAPM+'s approach of using predicted potential outcomes for matching decisions. Tabord-Meehan (2023) proposes stratification trees for two-stage trials that use first-wave data to determine optimal stratification schemes through decision tree algorithms. This method differs from BAPM+ in that it focuses on a class of stratification trees with varying assignment probabilities across strata. Li & Ding (2020) examine the theoretical properties of combining rerandomization with linear regression adjustment and similarly recognize that design and analysis choices are complementary rather than substitutable. However, their approach uses predetermined balance criteria in single-stage designs and linear regression adjustment, while BAPM+ learns to improve stratification adaptively across batches and uses a more flexible regression adjustment estimator.

The remainder of this paper is organized as follows. Section 2 establishes the setup for the framework. Section 3 presents the batch-adaptive pair-matching algorithm (BAPM), including an overview of the approach, a detailed description of the algorithm, and the distance functions used for pair matching. Section 4 discusses inference: Section 4.1 reviews an estimator for asymptotically exact inference without covariate adjustment, and Section 4.2 introduces a robust regression adjustment estimator. In Section 5, I analyze the relationship between stratification and regression adjustment and discuss the conditions under which researchers should stratify even when they plan to use covariate adjustment in their analysis. In Section 6, I evaluate the performance of BAPM with robust regression adjustment (BAPM+) relative to common alternatives with simulations that use both synthetic data and more realistic data derived from a political science experiment. Section 7 concludes with practical recommendations for social scientists using randomized experiments.

## 2 The Setup

I follow the potential outcomes framework in the Neyman-Rubin causal model (Neyman 1923, Rubin 1974). Let $i \in \{1, ..., 2N\}$ be the index of a participant in an experiment with a sample size of $2N$, randomly drawn from a superpopulation of interest. Let $Y_i \in \mathbb{R}$ be the observed outcome of interest for participant $i$, $Z_i \in \{0, 1\}$ be an indicator of whether participant $i$ is assigned to the treatment ($Z_i = 1$) or control ($Z_i = 0$) condition, and $X_i \in \mathbb{R}^k$ be the vector of observable covariates for participant $i$. I assume these covariates are observed *prior* to the start of the experiment. Let $Y_i(1)$ and $Y_i(0)$ be the potential outcomes for participant $i$ under treatment and control, respectively.[1] The potential outcomes map onto the observed outcome by the following formula:

$$Y_i = Y_i(1)Z_i + Y_i(0)(1 - Z_i).$$

The target causal estimand is the population-level average treatment effect (ATE):

$$\tau = E[Y_i(1) - Y_i(0)].$$

With access to both potential outcomes for each unit, the optimal stratification scheme for minimizing the mean squared error (MSE) of the ATE estimator is to pair units based on the expected sum of potential outcomes for each unit, conditional on the observed covariates (Bai 2022). This theoretical result informs the design of the algorithm I introduce in this paper. I discuss this result in more detail in Appendix Section 9.1.

In practice, we do not have access to potential outcomes. When sequential treatment assignment and outcome collection are feasible, we can use data from the first batch to estimate the potential outcomes and then use these estimates to inform the pairing. For the fitted predictive model to be useful, we need to assume the data generation process is stable across the two batches.

Before introducing the algorithm, I make the following assumptions about the potential outcomes and their relationship with the covariates.

---

[1] Although I focus on the binary treatment case for simplicity, researchers can combine BAPM with algorithms that seek to divide units into groups of more than two units, such as Karmakar (2022) and Higgins et al. (2016), to accommodate designs with multiple treatment arms.

**Assumption 1.** *(Regularity conditions)*

1. $0 < \mathbb{E}[\text{Var}(Y_i(z)|X_i)]$ *for* $z \in \{0, 1\}$.
2. $\mathbb{E}[Y_i^2(z)] < \infty$ *for* $z \in \{0, 1\}$.

**Assumption 2.** *(i.i.d.) Throughout the experiment,* $\forall i, j \in \{1, ..., 2N\}$ *and* $i \neq j$,

1. $(Y_i(1), Y_i(0), X_i) \perp (Y_j(1), Y_j(0), X_j)$ *and*
2. $(Y_i(1), Y_i(0), X_i) \stackrel{d}{=} (Y_j(1), Y_j(0), X_j)$.

This states that observations are independent and identically distributed (i.i.d.) throughout the experiment, a common assumption under the superpopulation framework (Imbens & Rubin 2015). Note that this assumption does not require the treatment status to be independent across units.

## 3 Batch-Adaptive Pair Matching (BAPM)

### 3.1 Overview

The Batch-Adaptive Pair Matching (BAPM) algorithm divides data into batches and uses data from earlier batches to inform pairing decisions in later batches. This method is distinct for its combination of adaptivity and cross-batch matching. For simplicity, I use the two-batch case to discuss the algorithm. The algorithm can be extended to cases with more than two batches.

The researcher plans to conduct an experiment on a sample of units, for which they have access to data on a range of covariates. To use the BAPM algorithm, the researcher begins by randomly splitting the sample into two batches. For the first batch, the researcher uses complete randomization to assign treatment and then collect data on realized outcomes. Then, the researcher fits models to this initial set of data to predict the potential outcomes of treatment and control for each participant, based on their individual covariates.

After the first round is concluded, a second batch of participants enters the study. Instead of immediately subjecting this new group to randomization and treatment, the researcher applies the predictive models developed from the first batch to the second batch. This allows the researcher to predict the potential outcomes. Following this, the researcher matches each participant with another one in the entire pool of participants. This matching uses the predicted potential outcomes for units in both batches. The researcher then assigns the treatment status for those in the second batch (1) to the opposite of the treatment status of their matched units in the first batch if the matched unit is from the first batch or (2) with pairwise randomization if the matched unit is from the second batch.

This approach has important advantages compared to complete randomization or conventional stratification on only covariates. By using information about the outcome-covariate relationship in forming matched pairs, researchers can reduce the scope for outcome-consequential covariate imbalance and thus reduce variance. By running a separate stratified experiment informed by the first batch, BAPM allows the researcher to use this predictive information in a more efficient way than by simply following complete randomization.

It also improves upon within-batch blocking methods that do use outcome information. First, it allows for matched pairs to be formed across batches rather than restricting matches within each batch, expanding the pool of potential matches for each unit. Second, BAPM can rely more heavily on one potential outcome model than the other if the two models differ in predictive accuracy.

## 3.2 The Algorithm

BAPM is designed to pair units across the two batches after learning a predictive model using the first batch. This predictive model maps the covariates onto estimates of the potential outcomes. The algorithm (Algorithm 1) begins by taking a set of covariates $X$ for $2N$ units and a procedure for estimating outcome models $f_1$ and $f_0$. The researcher should precommit to such a training procedure to avoid overfitting and the bias that may arise out of it if the goal is to pair units across batches.

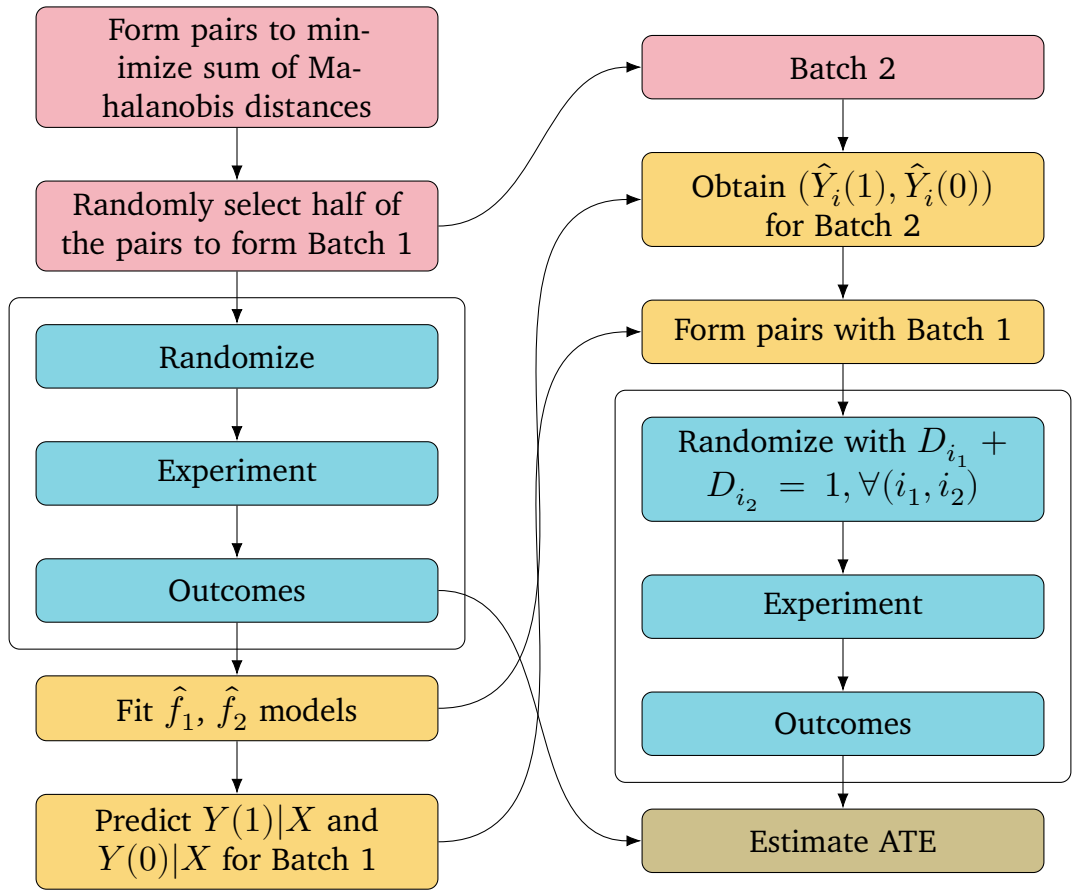

Figure 1: Batch-Adaptive Pair Matching

The algorithm then proceeds as follows. First, it finds the pairing scheme $\tilde{\lambda}$ that minimizes the sum of pairwise Mahalanobis distances for the *entire* set of units. It then uses pairwise randomization to assign treatment status to the units in the first batch. After the experiment for the first batch is conducted and the outcomes are collected, the algorithm fits potential outcome models $\hat{f}_0$ and $\hat{f}_1$ for $Y(0)$ and $Y(1)$ based on the data from the first batch.

These models are used to predict potential outcomes for units in both batches. For treated units in the first batch and all units in the second batch, the algorithm predicts their control outcomes using an outcome model fitted on all control units in the first batch. Similarly, for control units in the first batch and all units in the second batch, it predicts treatment outcomes using an outcome model fitted on all treatment units in the first batch. To predict treatment outcomes for treated units and control outcomes for control units in the first batch, the algorithm uses a leave-one-out strategy. For each treated (or control) unit, it excludes that unit to fit a new model for the treatment (or control) condition and then predicts the excluded unit's outcome under treatment (or control).

With these predictions, the algorithm calculates a pairwise distance $d_{i,j}(\lambda)$ for each possible pair of units across both batches. For units from the first batch, it imposes a constraint that pairs must consist of one treated and one control unit by setting the distance to infinity for pairs of first-batch units that do not meet this criterion. The algorithm then seeks the pairing scheme $\hat{\lambda}$ that minimizes the sum of pairwise distances across all pairs. The optimization problem for determining the pairing scheme is thus

$$\hat{\lambda} = \arg\min_{\lambda} \sum_{i,j \in B_1 \cup B_2} d_{i,j}(\lambda)$$

where $d_{i,j}$ is a distance metric that accounts for the predictive relationship between the covariates and the outcome. In practice, we can use the blossom algorithm to solve this optimization problem.

**Algorithm 1** BAPM

**Require:** Covariates $X \in \mathbb{R}^{2N \times k}$ for $2N$ units
1: Find pairing scheme $\tilde{\lambda}$ such that $\tilde{\lambda} = \arg\min_\lambda \sum_{i \neq j \in B_1 \cup B_2} d^M_{i,j}(\lambda)$
2: **for** each pair $(i,j) \in \hat{\lambda}_M$ **do**
3:     Randomly assign $Z_i \in \{0,1\}$ and set $Z_j = 1 - Z_i$
4: **end for**
5: Split $\tilde{\lambda}$ into two batches $\hat{\lambda}_1^M$ and $\hat{\lambda}_2^M$, with $|\hat{\lambda}_1^M| = N_1$ and $|\hat{\lambda}_2^M| = N_2$. Let $B_1 = \{i | i \in k, k \in \hat{\lambda}_1^M\}$ and $B_2 = \{i | i \in k, k \in \hat{\lambda}_2^M\}$.
6: Conduct the experiment for $B_1$ and collect outcome data $\mathbf{Y}_{B_1}$.
7: Fit models $\hat{f}_0$ for $Y(0)$ and $\hat{f}_1$ for $Y(1)$ based on $(\mathbf{D}_{B_1}, X_{B_1}, \mathbf{Y}_{B_1})$
8: **for** $i \in B_1^{(1)} \cup B_2$ and $j \in B_1^{(0)} \cup B_2$ **do**
9:     Predict $Y_i(0) | X_{i_0}$: $\hat{Y}_i(0) = \hat{f}_0(X_{i_0})$ and
10:     Predict $Y_{j_1}(1) | X_{j_1}$: $\hat{Y}_{j_1}(1) = \hat{f}_1(X_{j_1})$
11: **end for**
12: **for** each $i_1$ in the treated group in $B_1$ **do**
13:     Create a subset of $B_1$ excluding $i_1$: $B_1^{(1)} \setminus \{i_1\} = \{i \in B_1 | Z_i = 1, i \neq i_1\}$
14:     Fit model $\hat{f}_1$ for $Y(1)$ based on $(\mathbf{D}_{B_1^{(1)} \setminus \{i_1\}}, X_{B_1^{(1)} \setminus \{i_1\}}, \mathbf{Y}_{B_1^{(1)} \setminus \{i_1\}})$
15:     Predict $Y_{i_1}(1) | X_{i_1}$: $\hat{Y}_{i_1}(1) = \hat{f}_1(X_{i_1})$
16: **end for**
17: **for** each $i_0$ in the control group in $B_1$ **do**
18:     Create a subset of $B_1$ excluding $i_0$: $B_1^{(0)} \setminus \{i_0\} = \{i \in B_1 | Z_i = 0, i \neq i_0\}$
19:     Fit model $\hat{f}_0$ for $Y(0)$ based on $(\mathbf{D}_{B_1^{(0)} \setminus \{i_0\}}, X_{B_1^{(0)} \setminus \{i_0\}}, \mathbf{Y}_{B_1^{(0)} \setminus \{i_0\}})$
20:     Predict $Y_{i_0}(0) | X_{i_0}$: $\hat{Y}_{i_0}(0) = \hat{f}_0(X_{i_0})$
21: **end for**
22: Calculate $d^w_{i,j}(\lambda)$ for all $i,j \in B_1 \cup B_2$, $i \neq j$
23: Set $d^w_{i,j} = \infty$ if $Z_i + Z_j \neq 1$ for all $i,j \in B_1$, $i \neq j$
24: Find pairing scheme $\tilde{\lambda}$ such that $\hat{\lambda} = \arg\min_\lambda \sum_{i \neq j \in B_1 \cup B_2} d^w_{i,j}(\lambda)$
25: Initialize empty lists for dependent pairs, $\hat{\lambda}^{\text{dep}}$, and independent pairs, $\hat{\lambda}^{\text{indep}}$
26: **for** each pair $(i,j)$ in $\hat{\lambda}$ **do**
27:     **if** $i,j \in B_2$ **then**
28:         Randomly assign $Z_i \in \{0,1\}$ and set $Z_i = z$, $Z_j = 1 - d$
29:     **end if**
30:     **if** $i \in B_1$ and $j \in B_2$ **then**
31:         Set $Z_j = 1 - Z_i$
32:     **end if**
33: **end for**
34: **for** each $j \in B_2$ **do**
35:     Conduct experiment and collect outcome data
36: **end for**

The algorithm allows for the possibility that the best match for a unit in the second batch may be in the first batch. This is a key advantage of the algorithm over within-batch pairing (Bai 2022).[2]

For pairs that consist of one unit from the first batch and one from the second, the treatment assignment for the unit from the second batch inherits the randomness used for assigning units in the first batch. Thus, for units in the second batch, even though we use information about the relationship between the treatment and the potential outcomes to match the units with those in the first batch, this information is not used in determining the treatment status for the second batch. This preserves the independence of treatment assignment from the potential outcomes, conditional on the covariates for a given unit.

## 3.3 The Distance Function

To calculate the distance matrix, researchers might be tempted to use a metric that weights the two predicted potential outcomes equally, such as the Euclidean distance. However, the predictive models for potential outcomes may differ in their accuracy. In the most extreme case, if one of the models is completely uninformative, adding it to the distance function is akin to adding an uninformative covariate. To guard against this, we can use the Mahalanobis distance for the two predicted potential outcomes, weighted by the predictive accuracy of the models. Write the potential outcome predictions for unit $i$ as $\hat{\mathbf{y}}_i = (\hat{Y}_i(1), \hat{Y}_i(0))^T$. The distance function $d_{i,j}^w(\lambda)$ for pairs $(i,j) \in \lambda$ is thus:

$$d_{i,j}^w(\lambda) = \sqrt{(\hat{\mathbf{y}}_i - \hat{\mathbf{y}}_j)^T W S^{-1} W (\hat{\mathbf{y}}_i - \hat{\mathbf{y}}_j)} \quad (1)$$

where $S$ is the sample covariance matrix of the predicted potential outcomes $S(\hat{\mathbf{Y}}) = \frac{1}{n-1}\sum_{i=1}^{n}(\hat{\mathbf{y}}_i - \overline{\hat{\mathbf{y}}})(\hat{\mathbf{y}}_i - \overline{\hat{\mathbf{y}}})^T$ and $W = \text{diag}(w)$ is the diagonal weight matrix. The vector of weights can be determined by some measure of predictive accuracy, such as the $R^2$ of the predictive models for the potential outcomes.

---

[2]Researchers may be concerned about forming new pairs using revealed data. However, for inference in the superpopulation framework, treatment assignment can be dependent as long as a unit's treatment status is independent of the potential outcomes. Such dependence is common in empirical practice. For example, when researchers use complete randomization, it induces dependence in treatment status across all units.

# 4 Inference

For inference, I first extend the non-covariate-adjusted $t$-test proposed by Bai et al. (2022, henceforth BRS $t$-test) to BAPM. I then introduce a robust regression adjustment estimator that can further improve efficiency.

## 4.1 Inference without Covariate Adjustment

For the main results in Bai et al. (2022)[3] to be valid, two assumptions need to hold for the matched pairs: (1) units in each pair are "close" in their covariate space and (2) units in adjacent pairs are also "close" in the covariate space. Provided assumption (1) is satisfied, we can often find a permutation of the resulting pairs that satisfies assumption (2). Bai (2022) extends the results by showing that the estimator is still valid when within-pair units and units in adjacent pairs are close in transformations of the covariate space, provided that these transformations meet mild regularity conditions laid out in Assumption 3 in their paper. I restate the assumptions in Appendix Section 9.2. Here, I show that the estimator is also valid for BAPM under mild conditions.

Note that Bai et al. (2022) assume pairwise randomization, where, within each pair, the treatment status is uniformly distributed over $(0, 1), (1, 0)$, and pairwise independence, where the pairwise treatment status $(Z_i, Z_j)$ for each pair $(i, j)$ is i.i.d. While these assumptions are needed for the theoretical results on randomization inference in their paper, they are not necessary for the results on superpopulation inference.[4]

The main result in Bai et al. (2022) on the distribution of the difference-in-means esti-

---
[3]Bai et al. (2022) developed these methods for matched pair designs, with stratification either agnostic to the covariate-outcome relationship (Bai et al. 2022) or occurring within batches (Bai 2022).
[4]Interested readers can refer to the proof for Lemma S.1.4 in the Appendix of Bai et al. (2022).

mator thus holds:

$$\sqrt{N}(\hat{\tau} - \tau_0) \xrightarrow{d} \mathcal{N}(0, v_0),$$

where $\hat{\tau}$ is the difference-in-means estimator, $\tau_0$ is the true ATE, and

$$v_0 = \text{var}[Y_i(1)] + \text{var}[Y_i(0)] - \frac{1}{2}\mathbb{E}\left[(\mathbb{E}[Y_i(1)|X_i] - \mathbb{E}[Y_i(1)]) + (\mathbb{E}[Y_i(0)|X_i] - \mathbb{E}[Y_i(0)])\right]^2. \quad (2)$$

To apply the BRS $t$-test, the pairing scheme $\hat{\lambda}$ needs to satisfy the following assumption:

**Assumption 3.** *The pairing satisfies:*

$$\frac{1}{n} \sum_{1 \leq j \leq n} |X_{\pi(2j)} - X_{\pi(2j-1)}|^r \xrightarrow{p} 0$$

*for $r = 1$ and $r = 2$.*

Bai et al. (2022) show that a pairing scheme that minimizes the sum of pairwise distance satisfies this assumption when there is one covariate. When there are multiple covariates, they showed that the minimization of the sum of pairwise distances gives a bound on this sum and that this pairing scheme satisfies Assumption 3 when mild conditions hold. For the pairing produced by the BAPM algorithm to satisfy Assumption 3 (conditional on the pairing scheme that minimizes the sum of pairwise distance satisfying Assumption 3), we need to make an additional assumption about the predictive model.

**Assumption 4.** *The predictive model for the sum of potential outcomes, $F(X)$, is Lipschitz continuous.*

**Proposition 1.** *Given a pairing scheme $\hat{\lambda}$ that satisfies Assumption 3, the pairing scheme $\hat{\lambda}$ produced by the BAPM algorithm satisfies Assumption 3 when it satisfies Assumption 4.*

See Section 9.3 for the proof. For $2N$ units, the BRS $t$-test statistic is given by

$$T_N^{adj} = \frac{\sqrt{N}(\hat{\tau} - \tau_0)}{\hat{v}_N} \qquad (3)$$

where

$$\hat{v}_N = \hat{\delta}_N^2 - \frac{1}{2}\left(\hat{\lambda}_N^2 + \hat{\tau}_N\right)$$
$$\hat{\delta}_N^2 = \frac{1}{N(N-1)} \sum_{1 \leq m \leq N} (Y_\pi(2m) - Y_\pi(2m-1))^2,$$
$$\hat{\lambda}_N^2 = \frac{2}{N} \sum_{1 \leq m \leq \lfloor N/2 \rfloor} ((Y_\pi(4m-3) - Y_\pi(4m-2))(Y_\pi(4m-1) - Y_\pi(4m)))$$
$$= \times (D_\pi(4m-3) - D_\pi(4m-2))(D_\pi(4m-1) - D_\pi(4m))$$

$\hat{\delta}_N^2$ is the sample variance of the pairwise differences in the observed outcomes. $\hat{\lambda}_N^2$ first arranges the matched pairs into adjacent "pairs of pairs" and then, for each new pair of pairs, averages over products of the two pairwise differences in the observed outcomes.

## 4.2 Inference with Covariate Adjustment

After the experiment concludes, we have access to data on the realized outcomes and treatment status for units in the second batch. This gives us additional information about the relationship between the covariates and the potential outcomes. The BRS $t$-test is asymptotically exact, but it does not use this information in the analysis. To further improve the efficiency of the ATE estimator, we can use robust regression adjustment based on the Neyman orthogonal score (Robins & Rotnitzky 1994, Chernozhukov et al. 2018). I propose a variant that accounts for the stratified structure of the data. To use this method, we first pair the pairs, like we did for the BRS $t$-test above, to form blocks of four units. We then use a stratified estimator with stratum-specific regression adjustment. The estimator is given by

$$\widehat{\tau} = \frac{1}{N} \sum_{b=1}^{B} n_b \widehat{\tau}_b$$

with

$$\widehat{\tau}_b = \frac{1}{n_b} \sum_{i \in S_b} \left\{ \frac{Z_i}{\pi_b}[Y_i - \widehat{\mu}_1^{-b}(X_i)] - \frac{1-Z_i}{1-\pi_b}[Y_i - \widehat{\mu}_0^{-b}(X_i)] + \widehat{\mu}_1^{-b}(X_i) - \widehat{\mu}_0^{-b}(X_i) \right\} \quad (4)$$

where $S_b$ is the set of units in block $b$, $n_b = |S_b|$, $\pi_b = \frac{\sum_{i \in S_b} Z_i}{n_b}$ is the proportion of treated units in block $b$, and $\widehat{\mu}_1^{(-b)}(X_i)$ and $\widehat{\mu}_0^{(-b)}(X_i)$ are the predicted potential outcomes for the treated and control units, respectively, obtained by models fitted on a subset of the data that does not include units in block $b$.

The Neyman-style variance estimator is

$$\widehat{V}(\widehat{\tau}) = \frac{1}{N^2} \sum_{b=1}^{B} n_b^2 \left( \frac{\widehat{\sigma}_{1b}^2}{n_{1b}} + \frac{\widehat{\sigma}_{0b}^2}{n_{0b}} \right) \quad (5)$$

where

$$\widehat{\sigma}_{zb}^2 = \frac{1}{n_{zb} - 1} \sum_{i \in S_b} \left\{ \frac{Z_i}{\pi_b}[Y_i - \widehat{\mu}_z^{-b}(X_i)] + \widehat{\mu}_z^{-b}(X_i) \right\}^2$$

for $z \in \{0, 1\}$.

## 5 Do We Still Need to Stratify If We Plan to Use Covariate Adjustment?

Researchers might wonder whether they can still get efficiency improvements from stratification if they plan to use regression adjustment. The answer is yes: stratification makes the covariate adjustment problem milder, which leads to better estimates when the estimated model for regression adjustment is imperfect compared the oracle one (as is always the case). I explain this more formally below by expressing the distribution of the regression-adjusted estimator in terms of covariate imbalance and modeling error and showing how reducing imbalance produces better estimates in the presence of modeling error. The error can approach zero if we have either or both of the following: exact balance or the true regression model. In finite samples the latter does not hold, so design choices that shrink imbalance still drive down the error even with regression adjustment.

## 5.1 Formal Exposition

This section formalizes the claims above. Let $X_i \in \mathbb{R}^d$ be the feature vector that enters the regression adjustment. Suppose we fit two separate prediction models characterized by the following coefficients: $\hat{\beta}_1$ from regressing the observed outcomes $Y_j$ on $X_j$ using only the treated units $j$ that lie outside the fold containing unit $i$ and $\hat{\beta}_0$ with the control observations in the complementary folds. Consider the following regression-adjusted estimator for the ATE:

$$\hat{\tau} = \frac{1}{n} \sum_{i=1}^{n} \left\{ \left[ Z_i \frac{Y_i}{\pi} - (1 - Z_i) \frac{Y_i}{1 - \pi} \right] - X_i^\top \left[ \frac{Z_i}{\pi} \hat{\beta}_1 - \frac{1 - Z_i}{1 - \pi} \hat{\beta}_0 \right] + X_i^\top \left( \hat{\beta}_1 - \hat{\beta}_0 \right) \right\}. \tag{6}$$

Let $\beta_z^*$ be the following projection of the potential-outcome regression surface $\mu_z(X) = \mathbb{E}[Y_i(z) | X_i = X]$ onto the span of $X$:

$$\beta_z^* = \arg\min_{b \in \mathbb{R}^d} \mathbb{E}\left[ \{ \mu_z(X) - X^\top b \}^2 \right] \tag{7}$$

Let $\delta_{1n} = \hat{\beta}_1 - \beta_1^*$, $\delta_{0n} = \hat{\beta}_0 - \beta_0^*$, and $\Delta_n$ be the imbalance vector

$$\Delta_n = \frac{1}{n} \sum_{i=1}^{n} (Z_i - \pi) X_i. \tag{8}$$

Adding $-\tau$ to both sides of Equation 6 and re-arranging the terms, we have

$$\hat{\tau} - \tau = \sum_{i=1}^{n} \psi_i - \Delta_n^\top \Big(\frac{\hat{\beta}_1 - \beta_1^*}{\pi} + \frac{\hat{\beta}_0 - \beta_0^*}{1-\pi}\Big),$$

The error (or "plug-in bias") for this estimator can be written as (Hines et al. 2022):[5]

$$(\hat{\tau} - \tau) = \sum_{i=1}^{n} \psi_i - \Delta_n^\top \Big(\frac{\delta_{1n}}{\pi} + \frac{\delta_{0n}}{1-\pi}\Big). \tag{9}$$

where $\psi_i$ is the influence function for the regression-adjusted estimator:

$$\psi_i = \frac{Z_i(Y_i - X_i^\top \beta_1^*)}{\pi} - \frac{(1-Z_i)(Y_i - X_i^\top \beta_0^*)}{1-\pi} - \tau. \tag{10}$$

We can view $\psi_i$ as the "irreducible" variation in individual treatment effects. The second term in Equation 9 is an error due to imbalance between treatment and control groups in the projected covariate space $X$. It consists of two parts: the first is due to imbalance in the relevant covariate space $X$, and the second is due to the estimation of the learned coefficients $\hat{\beta}_1$ and $\hat{\beta}_0$. Write $\delta_n = \frac{\delta_{1n}}{\pi} + \frac{\delta_{0n}}{1-\pi}$ and $\Sigma = n^{-1} X^\top X$. The following proposition summarizes the expected squared error of the regression-adjusted estimator.

**Proposition 2.** *The expected squared error of the regression-adjusted estimator is*

$$\mathbb{E}[(\hat{\tau} - \tau)^2] = \frac{\text{Var}(\psi_i)}{n} + \frac{\pi(1-\pi)}{n} \mathbb{E}[\delta_n^\top \Sigma \delta_n]. \tag{11}$$

Now suppose the expermenter adopts an assignment mechanism such that balance is enforced on the span of a matrix $B$. Let $P_B$ be the projector onto this span and set $P_B^\perp = I - P_B$. Exact balance implies $P_B \Delta_n = 0$. Decompose $\delta_n$ as $\delta_n = \delta_{n,B} + \delta_{n,R}$ with $\delta_{n,B} = P_B \delta_n$. Because $(P_B \Delta_n)^\top \delta_{n,B} = 0$, we have

$$\delta_n^\top \Sigma \, \delta_n = (\Delta_n^\top \delta_{n,R})^2, \qquad \delta_{n,R} = P_B^\perp \delta_n.$$

---
[5]Note that, in the literature on double machine learning, there is often a remaining term that is due to the estimation of the propensity scores. This term is not present for randomized experiments where the propensity scores are known.

An analog to the second term in Equation 11 is thus

$$\begin{aligned}
\frac{\pi(1-\pi)}{n} \mathbb{E}[(\Delta_n^\top \delta_{n,R})^2] &= \frac{\pi(1-\pi)}{n} \mathbb{E}[\delta_{n,R}^\top \Sigma \, \delta_{n,R}] \\
&= \frac{\pi(1-\pi)}{n} (1-\eta_B^2) \mathbb{E}[\delta_n^\top \Sigma \, \delta_n]
\end{aligned} \quad (12)$$

where

$$\eta_B^2 = \frac{\|P_B(\beta_1^* - \beta_0^*)\|_\Sigma^2}{\|\beta_1^* - \beta_0^*\|_\Sigma^2} \in [0,1].$$

Compared with the second term in Equation 11, the term is now attenuated by $1 - \eta_B^2$. Stratification thus reduces the MSE when both of the following hold: (1) the learned or conjectured projection used for stratification does approximate its population target on a non-trivial way (that is, $\eta_B^2 > 0$)[6], and (2) the learned adjustment coefficients $\hat{\beta}_1$ and $\hat{\beta}_0$ are imperfect (that is, $\delta_n \neq 0$).

Both of these conditions are likely to hold in social science experiments with imperfectly predictive covariates and moderate sample sizes that prevent learning errors $\delta_n$ from approaching zero. The extent to which stratification improves upon regression adjustment in terms of the MSE rests on the magnitude of the deviation of the learned or conjectured projection from its population target and the "goodness" of the projector used for stratification. The BAPM algorithm makes it much more likely that the projection, even if worse than the coefficients learned at the end of the experiment, still improves upon a design without such stratification. It does so by improving the balance in the covariate spaces spanned by the learned coefficients $\hat{\beta}_1$ and $\hat{\beta}_0$.

---

[6]This is another way of stating the common advice that stratification can help improve efficiency when the covariates are predictive of the outcomes.

# 6 Simulations

I evaluate the performance of BAPM with the BRS $t$-test and robust regression adjustment in a series of simulation studies.

## 6.1 Methods Considered

I compare BAPM with regression adjustment (BAPM+) against six alternative methods that vary across four key design features. The simulation studies thus include the following seven methods:

- **Batch-Adaptive Pair Matching with Regression Adjustment (BAPM+)**: Cross-batch pairing using flexible machine learning models to predict potential outcomes, combined with post-design regression adjustment.

- **Batch-Adaptive Pair Matching without Regression Adjustment (BAPM)**: Cross-batch pairing using flexible machine learning models to predict potential outcomes, without regression adjustment.

- **Within-Batch Pair Matching (WBPM)**: Pairs units within batches using flexible models to predict potential outcomes.

- **Within-Batch Pair Matching with OLS (WBPM-OLS)**: Pairs units within batches using penalized OLS-based distance measures (Bai 2022).

- **Mahalanobis Matching (MH)**: Pairing using Mahalanobis distance on baseline covariates.

- **Complete Randomization (CR)**: Standard complete randomization without regression adjustment.

- **Complete Randomization with Regression (CR+)**: Complete randomization with post-design robust regression adjustment.[7]

---

[7]I do not include OLS-based regression adjustment (Lin 2013), popular in the social sciences (Gerber & Green 2012) because it can lead to significant undercoverage in superpopulation inference. I address this in a separate paper.

For methods with stratified randomization and no regression adjustment (BAPM, WBPM, WBPM-OLS, MH), I use the BRS $t$-test for inference. For BAPM+ and CR+, I use robust regression adjustment. For CR, I use the Neyman variance estimator.

Table 1 summarizes these methods across five design features:

- **S**tratified (S): The method uses stratified rather than complete randomization.
- **O**utcome-conscious (O): The method uses information about the relationship between covariates and outcomes in the assignment process.
- **F**lexible outcome model (F): The method used for learning the outcome model is flexible and does not assume a functional form.
- **C**ross-batch (C): The method allows pairing or matching units across different batches.
- **A**djustment with regression (A): The method uses robust regression adjustment in the analysis stage.

Table 1: Methods considered in simulation study

| Method | Stratified | Outcome-conscious | Flexible outcome model | Cross-batch | Regression adjustment | Summary |
| --- | --- | --- | --- | --- | --- | --- |
| BAPM+ | ✓ | ✓ | ✓ | ✓ | ✓ | SOFCA |
| BAPM | ✓ | ✓ | ✓ | ✓ | × | SOFC~~A~~ |
| WBPM | ✓ | ✓ | ✓ | × | × | SOF~~CA~~ |
| WBPM-OLS | ✓ | ✓ | × | × | × | SO~~FCA~~ |
| MH | ✓ | × | × | × | × | S~~OFCA~~ |
| CR | × | × | × | × | × | ~~SOFCA~~ |
| CR+ | × | × | × | × | ✓ | ~~SOFC~~A |

## 6.2 Simulations with Synthetic Data

I first use synthetic data to assess the performance of BAPM+ with post-design regression adjustment. I compare the performance of BAPM+ against seven alternative methods with their corresponding inference approaches.

### 6.2.1 Data Generation Process

Each simulation replicate uses $n = 96$ observations[8] with seven base covariates $(X_1, ..., X_7)$. $(X_1, X_2, X_3, X_7) \sim \mathcal{N}(0, I_4)$, $X_5 \sim \mathcal{U}(0, 1)$. Some covariates are correlated: $X_4 \sim \mathcal{N}(X_5, 1)$ and $X_6 \sim \mathcal{N}(X_3, 1)$.

To test how methods perform under different information conditions, I vary what covariates the "analyst" observes. The design matrix $X$ depends on the number of relevant predictors ($n_{\text{rel}}$) and irrelevant predictors ($n_{\text{irr}}$). When irrelevant covariates are included, they are drawn from $\mathcal{N}(0, 1)$ to introduce higher-variance noise:

- When $n_{\text{rel}} = 0$: The analyst only sees irrelevant noise variables.
- When $n_{\text{rel}} = 5$: The analyst observes the five outcome-predictive covariates $X_1, X_2, X_3, X_6, X_7$ plus any irrelevant covariates.
- When $n_{\text{rel}} > 5$: The analyst observes $X_2, X_3, X_4, X_6, X_7$ plus additional predictive covariates (each drawn from $\mathcal{N}(0, 1)$) with decreasing importance, plus any irrelevant covariates.

### 6.2.2 Outcome Models

I use nonlinear potential outcome functions with flexible relationships that are challenging to model without machine learning techniques:

- $Y(0) = 0.1 \sin(X_1) + 0.3 \exp(X_2) + 0.1 \exp(|X_3|) + 0.2 X_4 X_6$
- $Y(1) = 0.1 \sin(X_1) + 0.1 \exp(X_5) + 0.3 \exp(X_2) + 0.2 X_4 X_7$

---

[8]The sample size is 96 rather than 100 to ensure that we do not have to discard any pair in forming the "pairs of pairs" for the BRS $t$-test. This makes comparisons across algorithms more straightforward.

When $n_{\text{rel}} > 5$, additional covariates contribute to both potential outcomes with coefficients linearly decreasing from 0.05 to 0.01 based on the number of additional covariates.

### 6.2.3 Simulation Design

Each scenario is defined by $(n_{\text{rel}}, n_{\text{irr}})$. I run 5,000 iterations per scenario, estimating the average treatment effect, standard errors, and 95% confidence interval coverage. The main results focus on $(n_{\text{rel}}, n_{\text{irr}}) \in \{(10, 10), (0, 20)\}$, which allow us to assess performance when the analyst has access to a mix of relevant and irrelevant covariates versus only irrelevant variables.

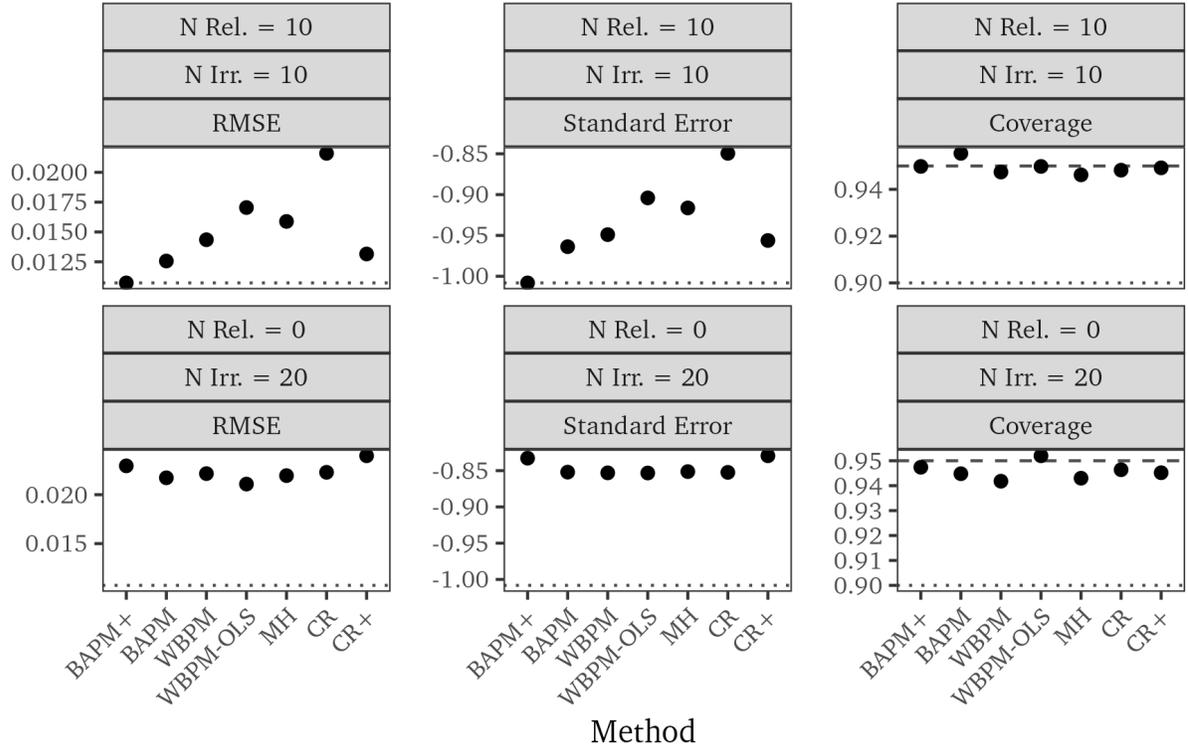

Figure 2: Results for synthetic data simulations with $n = 96$. The plot shows RMSE, standard errors, and coverage rates for the seven methods defined in Table 1. BAPM+: Batch-Adaptive Pair Matching with Regression Adjustment; BAPM: Batch-Adaptive Pair Matching; WBPM: Within-Batch Pair Matching; WBPM-OLS: Within-Batch Pair Matching with OLS; MH: Mahalanobis Matching; CR: Complete Randomization; CR+: Complete Randomization with Regression Adjustment. Results are presented for $(n_{\text{rel}}, n_{\text{irr}}) \in \{(10, 10), (0, 20)\}$. Dashed and dotted lines in the coverage panel indicate 95% and 90% nominal coverage rates, respectively; dotted lines in the RMSE and standard error panels indicate the value for BAPM+ with $n_{\text{rel}} = 10$ and $n_{\text{irr}} = 10$.

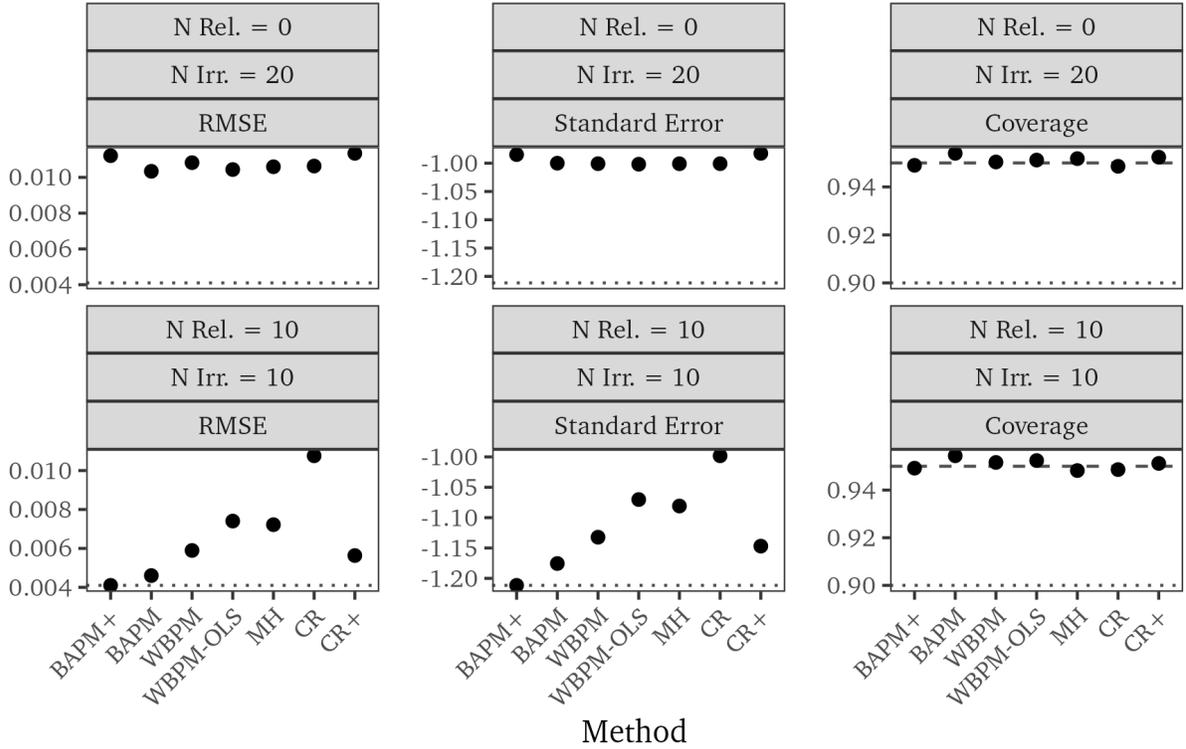

Figure 3: Results for synthetic data simulations with $n = 200$. The plot shows RMSE, standard errors, and coverage rates for the seven methods defined in Table 1. BAPM+: Batch-Adaptive Pair Matching with Regression Adjustment; BAPM: Batch-Adaptive Pair Matching; WBPM: Within-Batch Pair Matching; WBPM-OLS: Within-Batch Pair Matching with OLS; MH: Mahalanobis Matching; CR: Complete Randomization; CR+: Complete Randomization with Regression Adjustment. Results are presented for $(n_{\text{rel}}, n_{\text{irr}}) \in \{(10, 10), (0, 20)\}$. Dashed and dotted lines in the coverage panel indicate 95% and 90% nominal coverage rates, respectively; dotted lines in the RMSE and standard error panels indicate the value for BAPM+ with $n_{\text{rel}} = 10$ and $n_{\text{irr}} = 10$.

Figure 2 and Figure 3 present the simulation results for sample sizes of 96 and 200, respectively. Both figures display three key performance metrics: RMSE, standard errors, and coverage rates across different covariate scenarios.

When outcome-predictive covariates are available ($n_{\text{rel}} = 10, n_{\text{irr}} = 10$), BAPM+ achieves the lowest RMSE and standard errors, followed by BAPM, then the within-batch methods (WBPM and WBPM-OLS), Mahalanobis matching (MH), and complete randomization methods (CR and CR+). For $n = 200$, MH performs better than WBPM-OLS in terms of RMSE and standard errors, suggesting that when sample sizes are larger, Mahalanobis distance matching can outperform within-batch approaches that use less flexible outcome models.

When only irrelevant covariates are available ($n_{\text{rel}} = 0, n_{\text{irr}} = 20$), the methods perform roughly equally in terms of RMSE and average standard errors. This is expected: without outcome-predictive information, stratification methods cannot improve upon complete randomization in terms of efficiency. The similarity in performance across methods in this scenario validates that, on average, the adaptive methods do not perform worse than non-adaptive alternatives when the covariates are uninformative.[9]

For $n = 200$, all methods achieve nominal 95% coverage rates in both scenarios. For $n = 96$, coverage rates are generally close to the nominal level when outcome-predictive covariates are available, though some methods show slight undercoverage. When only irrelevant covariates are available ($n_{\text{rel}} = 0, n_{\text{irr}} = 20$), coverage rates for most methods dip below the nominal rate, with some falling below 90%. This undercoverage at smaller sample sizes with uninformative covariates suggests that inference can be more challenging when the sample size is limited and stratification cannot exploit outcome-relevant information.

## 6.3 Simulations with Semi-Realistic Data

To illustrate the practical performance of the BAPM algorithm with a realistic example, I constructed a semi-synthetic dataset based on experimental data from the CNN study by Broockman & Kalla (2025). The original authors conducted a randomized experiment to assess the impact of watching CNN on a range of attitudinal outcomes for Fox News viewers. For the purpose of the simulation, I focus on the thermometer ratings for Republican officials,

---

[9]Note that the worst-case MSE can be larger for stratified designs compared to complete randomization when the covariates are uninformative. If researchers are concerned about the worst-case MSE and seeks trade offs between that and average efficiency gains, they should use the design proposed by Harshaw et al. (2024).

which were measured on a scale from 0 to 100.

I preprocess the data by removing observations with missing values in the treatment or outcome variables and using iterative imputation to impute missing values for the remaining variables. I use the ensemble of gradient boosting models to predict counterfactual outcomes: for control units, I predicted potential outcomes under treatment as $\hat{Y}_1(0) = \frac{1}{M}\sum_{m=1}^{M} f_1^m(\mathbf{X}_0)$, and for treated units, I predict potential outcomes under control as $\hat{Y}_0(1) = \frac{1}{M}\sum_{m=1}^{M} f_0^m(\mathbf{X}_1)$. The final semi-synthetic dataset consists of 290 covariates $X$, the observed outcomes and the counterfactual outcomes for a "population" of 727 units. At each iteration, I randomly sample 96 units from the population without replacement.

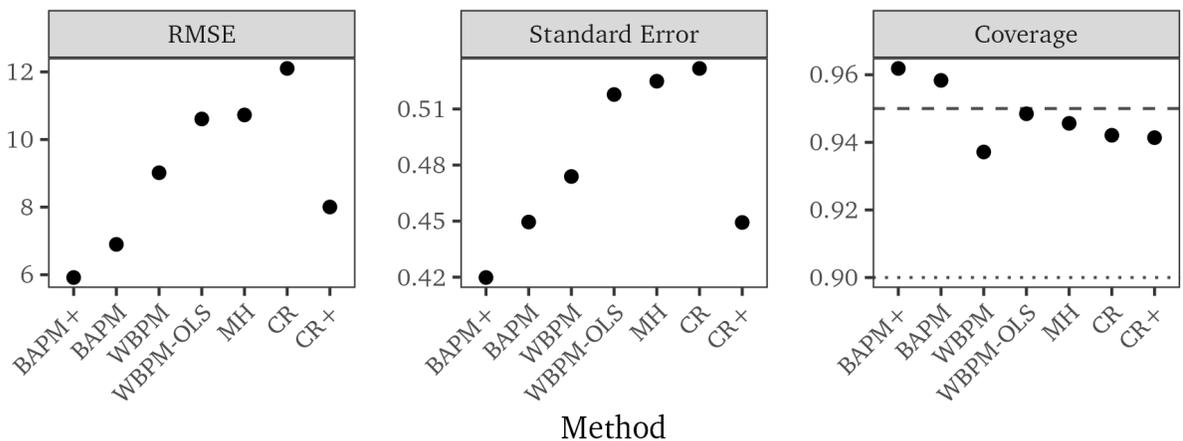

Figure 4: Results for semi-synthetic data simulations with $n = 96$. The plot shows RMSE, standard errors, and coverage rates for the seven methods defined in Table 1 using semi-synthetic data based on the CNN study. BAPM+: Batch-Adaptive Pair Matching with Regression Adjustment; BAPM: Batch-Adaptive Pair Matching; WBPM: Within-Batch Pair Matching; WBPM-OLS: Within-Batch Pair Matching with OLS; MH: Mahalanobis Matching; CR: Complete Randomization; CR+: Complete Randomization with Regression Adjustment. Dashed and dotted lines in the coverage panel indicate 95% and 90% nominal coverage rates, respectively.

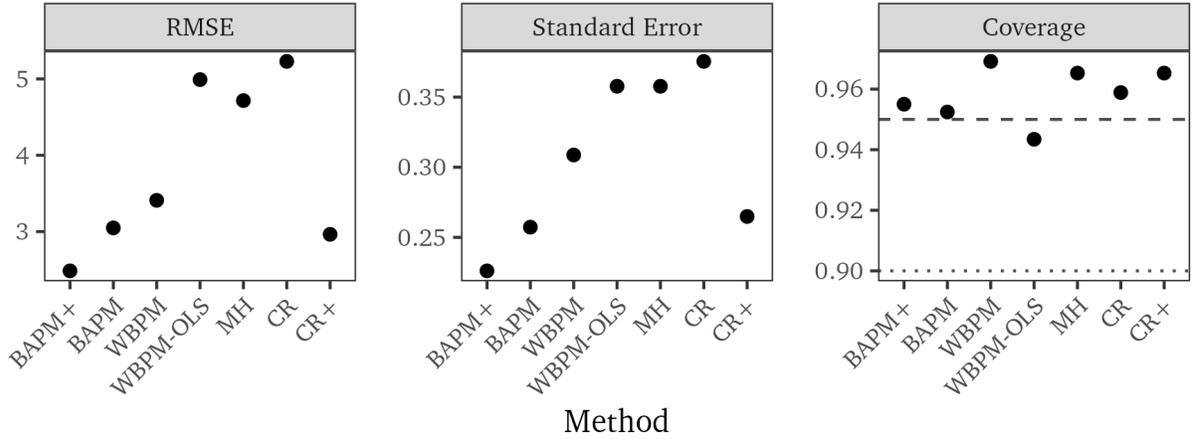

Figure 5: Results for semi-synthetic data simulations with $n = 200$. The plot shows RMSE, standard errors, and coverage rates for the seven methods defined in Table 1 using semi-synthetic data based on the CNN study. BAPM+: Batch-Adaptive Pair Matching with Regression Adjustment; BAPM: Batch-Adaptive Pair Matching; WBPM: Within-Batch Pair Matching; WBPM-OLS: Within-Batch Pair Matching with OLS; MH: Mahalanobis Matching; CR: Complete Randomization; CR+: Complete Randomization with Regression Adjustment. Dashed and dotted lines in the coverage panel indicate 95% and 90% nominal coverage rates, respectively.

Figure 4 and Figure 5 show the results from the semi-synthetic simulations based on the CNN study data for sample sizes of 96 and 200, respectively. These results are consistent with the patterns observed in the synthetic data simulations.

BAPM+ achieves the lowest RMSE and standard errors, followed by BAPM, then the within-batch methods, MH, and complete randomization approaches. Coverage rates show similar patterns to the synthetic simulations. For $n = 96$, MH and WBPM exhibit slight undercoverage, with rates occasionally dipping below the 90% threshold. These coverage issues are less pronounced at $n = 200$, where most methods achieve coverage rates above or close to the nominal 95% level.

Figure 4 shows that BAPM+ outperforms the other methods in terms of both the RMSE and the average standard error of the ATE estimator for N=96. The reduction in RMSE from using BAPM+ over only using post-design regression adjustment is about 24.4% of the true ATE. Relative to using BAPM without regression adjustment, it is about 11.9% of the true ATE. These results suggest the gain from using BAPM+ can be substantial in realistic settings.

Figure 5 shows the results for N=200. The patterns are similar to those for N=96, with BAPM+ achieving the lowest RMSE and standard errors. The reduction in RMSE from using BAPM+ over only using post-design regression adjustment is about 8.9% of the true ATE. Relative to using BAPM without regression adjustment, it is about 10.4% of the true ATE. The gains from adaptive stratification, compared to regression adjustment alone, are smaller than those for N=96, as expected because the larger sample size allows the learning error in regression adjustment to shrink.

# 7  Practical Recommendations

This paper presents a new stratification algorithm for randomized experiments that incorporates the relationship between the covariates and the potential outcomes in the pairing process. The method, BAPM+, can improve efficiency when covariate data is available ex ante and sequential experimentation is feasible. BAPM+ obtains a predictive model for the potential outcomes using data from earlier batches and pairs units across batches based on the predicted potential outcomes. It then uses robust regression adjustment in the analysis stage to gain further precision and efficiency. Through simulations, I demonstrate that this approach improves precision in sequential experimental settings relative to several contenders. I also show that the gains in MSE can be translated into practical gains in efficiency of the regression-adjusted ATE. The coverage rates of the CIs obtained using this framework are close to the nominal level.

Based on the analytical and simulation findings in this paper, I offer the following practical recommendations for researchers conducting randomized experiments in the social sciences. First, researchers should stratify whenever feasible, as the efficiency gains complement rather than substitute those from post-design adjustment in common scenarios.

BAPM+ is an outcome-conscious stratification method that can be helpful when researchers have access to comprehensive covariate data before the experiment begins and can collect outcome data from pilot studies or earlier experimental batches. In such sequential experimental settings, BAPM+ leverages the predictive relationship between covariates and outcomes that traditional stratification methods ignore, leading to substantial improvements in precision and efficiency.

Regardless of the stratification approach used, I recommend robust regression adjustment in the analysis stage. The simulations demonstrate that combining stratification with post-design adjustment yields the best performance across diverse scenarios. Even when stratification provides only modest gains on its own, the combination with regression adjustment consistently outperforms either approach used in isolation in terms of the precision and efficiency of the ATE estimates.

Stratification may be unnecessary only in the rare circumstances where researchers are highly confident that adjustment models will almost perfectly approximate the oracle ones.[10] However, this scenario is uncommon in social science applications, where sample sizes are often modest. Given the practical constraints facing most social science experiments, the insurance provided by good stratification remains valuable even when researchers plan to use regression adjustment.

---

[10]Zhang (2025) used a similar but distinct approach (with different notation) and arrived at an ostensibly different conclusion. However, the difference is in the emphasis: Zhang (2025) focused on the fact that regression adjustment can be used to achieve *some* of the gains from stratification.

justments in randomized experiments', *Proceedings of the National Academy of Sciences* **113**(45), 12673–12678.

Zhang, Z. (2025), 'A connection between covariate adjustment and stratified randomization in randomized clinical trials', *Statistical Methods in Medical Research* **34**(4), 829–844.

# 9  Appendix

## 9.1  Oracle Stratification

Let $g$ be a function that maps the covariates onto values on a real line: $g(x)\colon \mathbb{R}^k \to \mathbb{R}$. Bai (2022) shows that the optimal stratification should be based on the following function:

$$g^*(x) = \mathbb{E}\left[Y_i(1) + Y_i(0) | X_i = x\right].$$

Following their notation, I denote $\pi^g$ as a permutation of the units $\{1, ..., 2N\}$ such that $g_{\pi(r(1))} \leq \cdots \leq g_{\pi(r(2N))}$ and a pairing function for an experiment with $2N$ observations as $\lambda^g(X^n)$:

$$\lambda^g(X^n) = \left\{(\pi^g(2s-1), \pi^g(2s)) : 1 \leq s \leq N\right\}.$$

The oracle stratification strategy that minimizes the MSE involves creating pairs $(\pi^g(2s-1), \pi^g(2s))$ for $s$ ranging from 1 to $N$, where $\pi^g$ ranks units based on their values of $g$.

In practice, however, we do not have access to $Y_i(1) + Y_i(0)$ for any unit $i$. A workaround is to plug in the predicted potential outcomes $\hat{Y}_i(1)$ and $\hat{Y}_i(0)$ from predictive models for the potential outcomes. But this approach implicitly weights the predicted potential outcomes under treatment and control equally. This can lead to suboptimal performance if the predictive models for the potential outcomes have different predictive power. Consider, for example, an extreme case where the predictive model for $Y_i(1)$ is highly accurate while the predictive model for $Y_i(0)$ is completely off. In this case, stratifying based on $\hat{Y}_i(1) + \hat{Y}_i(0)$ would be strictly worse than stratifying based on $\hat{Y}_i(1)$ alone.

## 9.2 Assumptions for Exact Inference without Covariate Adjustment

I restate the assumptions for the BRS $t$-test (Bai et al. 2022). The first assumption is a smoothness condition on the conditional expectations of the potential outcomes.

**Assumption 5.** 1. $\mathbb{E}[Y_i(z)|X_i = x]$ and $\mathbb{E}[Y_i^2(z)|X_i = x]$ are Lipschitz for $z \in \{0, 1\}$.

The next two assumptions are on the pairing procedure.

**Assumption 6.** *The pairing satisfy:*

$$\frac{1}{n} \sum_{1 \leq j \leq n} |X_{\pi(2j)} - X_{\pi(2j-1)}|^r \xrightarrow{p} 0$$

*for $r = 1$ and $r = 2$.*

Assumption 6 states that the distance between the covariates of paired units is small.

**Assumption 7.** *The pairing satisfy:*

$$\frac{1}{n} \sum_{1 \leq j \leq \lfloor n/2 \rfloor} |X_{\pi(4j-k)} - X_{\pi(4j-\ell)}|^2 \xrightarrow{p} 0$$

*for any $k \in \{2, 3\}$ and $\ell \in \{0, 1\}$.*

Assumption 7 states that the distance between the covariates of paired pairs is small. If Assumption 3 holds, we can rank the pairs in such a way that Assumption 7 holds. Both conditions hold approximately if we find a pairing that minimizes the sum of pairwise distances between the covariates of paired units and rerank the pairs in such a way that adjacent pairs are close to each other in the ranking (Bai et al. 2022). We can find such a pairing with the blossom algorithm (Edmonds 1965).

## 9.3 Proof of Proposition 1

Given a predictive model $F(X)$ that is Lipschitz continuous, we have

$$\left|F(X_i) - F(X_j)\right| \leq L \left|X_i - X_j\right|$$

where $L$ is the Lipschitz constant.

This means, under the first pairing scheme $\tilde{\lambda}$, we have

$$\sum_{i \neq j \in B_1 \cup B_2} \left|F(X_{\hat{\pi}(2i-1)}) - F(X_{\hat{\pi}(2i)})\right| \leq L \times \sum_{i \neq j \in B_1 \cup B_2} \left|X_{\hat{\pi}(2i-1)} - X_{\hat{\pi}(2i)}\right|$$

With Assumption 3, we have

$$\sum_{i \neq j \in B_1 \cup B_2} \left|F(X_{\hat{\pi}(2i-1)}) - F(X_{\hat{\pi}(2i)})\right|^r \xrightarrow{p} 0$$

for $r = 1, 2$.

Under the second pairing scheme $\hat{\lambda}$, we have

$$\sum_{i \neq j \in B_1 \cup B_2} \left|F(X_{\hat{\pi}(2i-1)}) - F(X_{\hat{\pi}(2i)})\right| \leq \sum_{i \neq j \in B_1 \cup B_2} \left|F(X_{\tilde{\pi}(2i-1)}) - F(X_{\tilde{\pi}(2i)})\right| \quad (13)$$

This holds because the second pairing scheme $\hat{\lambda}$ minimizes

$$\sum_{i \neq j \in B_1 \cup B_2} \left|F(X_{\pi(2i-1)}) - F(X_{\pi(2i)})\right| \quad (14)$$

subject to the constraint

$$(\pi(2i-1), \pi(2i)) \notin \hat{\lambda} \text{ if } 2i-1, 2i \in B_1 \text{ and } D_{2i-1} = D_{2i},$$

which $\tilde{\lambda}$ also satisfies. In other words, if $\tilde{\lambda}$ minimizes Equation 14, then $\hat{\lambda} = \tilde{\lambda}$; otherwise, the inequality in Equation 13 holds strictly.

## 9.4 Proof of Proposition 2

Write

$$A_n = \frac{1}{n}\sum_{i=1}^n \psi_i, \qquad B_n = \Delta_n^\top\left(\frac{\delta_{1n}}{\pi} + \frac{\delta_{0n}}{1-\pi}\right) = \tilde{\Delta}_n^\top \delta_n$$

The estimator error is thus $\hat{\tau} - \tau = A_n - B_n$. Squaring and taking expectations gives

$$\mathbb{E}(\hat{\tau}) = \mathbb{E}[A_n^2] - 2\,\mathbb{E}[A_n B_n] + \mathbb{E}[B_n^2]. \tag{15}$$

First, I show that $\mathbb{E}[A_n] = 0$. In the main text, I defined

$$\psi_i = \frac{Z_i(Y_i - X_i^\top \beta_1^*)}{\pi} - \frac{(1-Z_i)(Y_i - X_i^\top \beta_0^*)}{1-\pi} - \tau.$$

Note

$$\mathbb{E}[Z_i|X_i, Y_i(0), Y_i(1)] = \pi, \qquad \mathbb{E}[1-Z_i|X_i, Y_i(0), Y_i(1)] = 1-\pi,$$

and, by definition of $\beta_z^*$, $\mathbb{E}[Y_i(z) - X_i^\top \beta_z^*|X_i] = 0$. Taking expectations with respect to the assignment mechanism therefore gives $\mathbb{E}[\psi_i|X_i] = 0$. Unconditional mean-zero follows by the law of iterated expectations. Because $\psi_i$ is mean-zero, the first term becomes

$$\mathbb{E}[A_n^2] = \frac{\mathrm{Var}(\psi_i)}{n}.$$

The second term is

$$\mathbb{E}[A_n B_n] = \frac{1}{n^2}\sum_{i=1}^n\sum_{j=1}^n \mathbb{E}\left[\psi_i\,(Z_j - \pi)\,X_j^\top \delta_n\right].$$

Conditional on $(X_i)_{i=1}^n$ and on the cross-fitted coefficients, $\mathbb{E}[(Z_j - \pi)\psi_i | X_1^n] = 0$ for every $i, j$, because $\psi_i$ is linear in $Z_i$ with coefficient zero mean and is independent of $Z_j$ for $j \neq i$. Hence
$$\mathbb{E}[A_n B_n] = 0.$$

For the third term, conditioning on $(X_i)_{i=1}^n$ and using independence of treatment draws, we have

$$\mathbb{E}[B_n^2 | (X_i)_{i=1}^n, \delta_n] = \pi(1-\pi)\, n^{-1}\, \delta_n^\top \Sigma\, \delta_n, \qquad \Sigma = \frac{1}{n}\sum_{i=1}^n X_i X_i^\top.$$

Unconditioning yields

$$\mathbb{E}[B_n^2] = \frac{\pi(1-\pi)}{n}\, \mathbb{E}[\delta_n^\top \Sigma\, \delta_n].$$

## 9.5 A Note on Attrition

Note that, while the BRS $t$-test depends on the paired structure of the data, the stratified regression adjustment estimator does not. We can thus still use stratified regression adjustment estimator even if there is attrition in the experiment, as long as the attrition is ignorable. In such cases, we can first restratify the remaining units into blocks of four using their predicted potential outcomes and then apply the regression adjustment within each block.

## 9.6 Alternatives to BAPM

To show the efficiency advantage of BAPM, I consider three plausible alternatives. The first is to use Mahalanobis distance for pair-matching across the entire sample without a batched design or adaptivity. Secondly, I use a batched matching approach that pairs units within batches. Units within the first batch will be matched using Mahalanobis distance. For the second batch, the matching process will utilize the predictive models trained on data from the first batch. The key difference between this approach and BAPM is that it conducts matching within each batch while BAPM allows for pairing across batches. Therefore, it incorporates information about the relationship between the covariates and the potential outcomes in the pairing for the second batch but does not use this information as optimally as BAPM. I call this approach Within-batch Pair Matching (WBPM). The third alternative is the method introduced by Bai (2022). It fits a predictive model based on data from a first batch or "small pilot" and pairs units in the second batch using a OLS-based distance function that penalizes uncertainty in the model. The key difference between this approach and WBPM is that it uses a penalized OLS distance function for pairing units in the second batch and does not use flexible machine learning methods for prediction. I call this approach Within-batch Pair Matching with Penalized OLS (WBPM-OLS).

### 9.6.1 Mahalanobis Pair Matching (MH)

A more flexible approach than discrete blocking uses distance measures, such as the Mahalanobis distance, to match units that are "close" in covariate space (Moore 2012). While this accommodates continuous variables and multiple covariates, it still cannot incorporate information about which covariates most strongly predict outcomes. Mahalanobis distance matching works well in practice when the number of covariates is small and information about the outcome is limited (Bai 2022). However, when information is available for both covariates and outcomes, this metric cannot accommodate the differential significance of covariates in predicting outcomes; it is still agnostic with respect to the predictive relationship between covariates and outcomes.

The first alternative approach minimizes the sum of pairwise Mahalanobis distances. Formally, the Mahalanobis distance between two observations with covariate vectors $\mathbf{x_1}, \mathbf{x_2}$ is

$$\mathbf{M}(\mathbf{x_1}, \mathbf{x_2}) = \sqrt{(\mathbf{x_1} - \mathbf{x_2})^T \Sigma^{-1} (\mathbf{x_1} - \mathbf{x_2})},$$

where $\Sigma$ represents the covariance matrix of the covariates.

The pairing scheme can thus be characterized as:

$$\hat{\lambda} = \arg\min_\lambda \sum_{i,j \in \{1,\ldots,2N\}, i \neq j} \mathbf{M}_{i,j}(\lambda). \tag{16}$$

### 9.6.2 Within-batch Pair Matching (WBPM)

For the first batch, units are paired using Mahalanobis pair matching, as in (16) but with $i, j \in B_1$. Then, after outcomes are drawn, I fit models to predict the potential outcomes using data from this first batch and use the models to predict the potential outcomes for units in the second batch. I then pair units in the second batch by minimizing the sum of pairwise distances between the sum of predicted potential outcomes:

$$\hat{\lambda}_{\text{WBPM}} = \arg\min_\lambda \sum_{i,j \in B_2, i \neq j} d_{i,j}^{WBPM}(\lambda),$$

where $d_{i,j}^{WBPM}(\lambda) = \sqrt{(\hat{\mathbf{y}}_i - \hat{\mathbf{y}}_j)^T S^{-1} (\hat{\mathbf{y}}_i - \hat{\mathbf{y}}_j)}$. $\hat{\mathbf{y}}_i$ and $\hat{\mathbf{y}}_j$ are the predicted potential outcomes for units $i$ and $j$, respectively, and $S$ is the covariance matrix of the predicted potential outcomes.

Inference can then be performed separately for the two batches and then aggregated.

### 9.6.3 Within-batch Pair Matching with Penalized OLS (WBPM-OLS)

This is the approach proposed by Bai (2022) for learning a predictive model for the potential outcomes using a pilot. The basic idea is to use data from an earlier batch to fit an ordinary least squares (OLS) model for estimating potential outcomes, then use this model to inform assignment in the main experiment (second batch). While this approach improves upon outcome-agnostic stratification, it has several shortcomings. First, pairing occurs only within batches, meaning that even when a unit's best match is in the other batch, it is paired with a suboptimal match in the same batch. The justification seems to be that once pairs for the first batch are formed, they cannot be broken up. However, pairing within batches is unnecessary for valid inference about treatment effects from a superpopulation perspective; it is thus unnecessarily restrictive and can be suboptimal. Second, OLS is not always the best predictive model for potential outcomes, especially in high-dimensional settings. Data-adaptive machine learning methods may offer substantial gains.

To maxmize comparability, I use Mahalanobis distance for pairing in the first batch, as in WBPM. For the second batch, I use the penalized OLS distance function that Bai (2022) proposed. The distance function is defined as

$$\hat{d}(\mathbf{x}_1, \mathbf{x}_2) = \left(\mathbf{x}_1'\hat{\beta} - \mathbf{x}_2'\hat{\beta}\right)^2 + (\mathbf{x}_1 - \mathbf{x}_2)'\hat{\Sigma}(\mathbf{x}_1 - \mathbf{x}_2).$$

where $\hat{\Sigma}$ is the covariance matrix of $\hat{\beta}$, the OLS coefficients, under homoskedasticity.

## 9.7 Additional Simulation Results

### 9.7.1 Simulations with Synthetic Data

Table 2 shows the results in tabular form for the simulations using synthetic data. T

### 9.7.2 Simulations with Semi-Realistic Data

This section presents additional simulation results using data based on the CNN study (Broockman & Kalla 2025). Table 4 and Table 5 show the results in tabular form for both sample sizes (96 and 200).

Table 2: Metrics of Simulations Using Nonlinear DGPs (Averaged over 5,000 iterations), N = 96. BAPM+: Batch-Adaptive Pair Matching with Regression Adjustment; BAPM: Batch-Adaptive Pair Matching; WBPM: Within-Batch Pair Matching; WBPM-OLS: Within-Batch Pair Matching with OLS; MH: Mahalanobis Matching; CR: Complete Randomization; CR+: Complete Randomization with Regression Adjustment.

| N Rel. | N Irr. | Method | ATE | SE | CI Length | Coverage | RMSE |
|---|---|---|---|---|---|---|---|
| **0** | **20** | **BAPM+** | **-0.107** | **0.145** | **0.567** | **0.948** | **0.151** |
| 0 | 20 | BAPM | -0.107 | 0.138 | 0.541 | 0.944 | 0.145 |
| 0 | 20 | WBPM | -0.110 | 0.138 | 0.539 | 0.942 | 0.146 |
| 0 | 20 | WBPM-OLS | -0.109 | 0.137 | 0.539 | 0.953 | 0.143 |
| 0 | 20 | MH | -0.109 | 0.138 | 0.542 | 0.945 | 0.144 |
| 0 | 20 | CR | -0.111 | 0.138 | 0.541 | 0.950 | 0.146 |
| 0 | 20 | CR+ | -0.111 | 0.146 | 0.572 | 0.947 | 0.152 |
| **10** | **10** | **BAPM+** | **-0.109** | **0.094** | **0.370** | **0.950** | **0.099** |
| 10 | 10 | BAPM | -0.110 | 0.105 | 0.412 | 0.955 | 0.109 |
| 10 | 10 | WBPM | -0.109 | 0.109 | 0.429 | 0.948 | 0.117 |
| 10 | 10 | WBPM-OLS | -0.108 | 0.122 | 0.477 | 0.950 | 0.127 |
| 10 | 10 | MH | -0.110 | 0.119 | 0.465 | 0.949 | 0.123 |
| 10 | 10 | CR | -0.108 | 0.139 | 0.545 | 0.949 | 0.145 |
| 10 | 10 | CR+ | -0.107 | 0.107 | 0.419 | 0.952 | 0.111 |

Table 3: Metrics of Simulations Using Nonlinear DGPs (Averaged over 5,000 iterations), N = 200. BAPM+: Batch-Adaptive Pair Matching with Regression Adjustment; BAPM: Batch-Adaptive Pair Matching; WBPM: Within-Batch Pair Matching; WBPM-OLS: Within-Batch Pair Matching with OLS; MH: Mahalanobis Matching; CR: Complete Randomization; CR+: Complete Randomization with Regression Adjustment.

| N Rel. | N Irr. | Method | ATE | SE | CI Length | Coverage | RMSE |
|---|---|---|---|---|---|---|---|
| **0** | **20** | **BAPM+** | **-0.109** | **0.102** | **0.400** | **0.945** | **0.107** |
| 0 | 20 | BAPM | -0.109 | 0.098 | 0.385 | 0.949 | 0.101 |
| 0 | 20 | WBPM | -0.110 | 0.098 | 0.384 | 0.949 | 0.102 |
| 0 | 20 | WBPM-OLS | -0.110 | 0.098 | 0.384 | 0.952 | 0.100 |
| 0 | 20 | MH | -0.111 | 0.098 | 0.385 | 0.947 | 0.102 |
| 0 | 20 | CR | -0.108 | 0.098 | 0.385 | 0.952 | 0.100 |
| 0 | 20 | CR+ | -0.108 | 0.103 | 0.402 | 0.954 | 0.104 |
| **10** | **10** | **BAPM+** | **-0.108** | **0.059** | **0.232** | **0.946** | **0.063** |
| 10 | 10 | BAPM | -0.107 | 0.065 | 0.253 | 0.954 | 0.067 |
| 10 | 10 | WBPM | -0.109 | 0.072 | 0.281 | 0.950 | 0.075 |
| 10 | 10 | WBPM-OLS | -0.110 | 0.083 | 0.325 | 0.951 | 0.085 |
| 10 | 10 | MH | -0.109 | 0.081 | 0.319 | 0.948 | 0.084 |
| 10 | 10 | CR | -0.111 | 0.099 | 0.387 | 0.950 | 0.102 |
| 10 | 10 | CR+ | -0.109 | 0.069 | 0.269 | 0.953 | 0.073 |

Table 4: Metrics of Simulations Using Data Based on the CNN Study (Averaged over 5,000 iterations), N = 96. BAPM+: Batch-Adaptive Pair Matching with Regression Adjustment; BAPM: Batch-Adaptive Pair Matching; WBPM: Within-Batch Pair Matching; WBPM-OLS: Within-Batch Pair Matching with OLS; MH: Mahalanobis Matching; CR: Complete Randomization; CR+: Complete Randomization with Regression Adjustment.

| Method | ATE | SE | CI Length | Coverage | RMSE |
| --- | --- | --- | --- | --- | --- |
| **BAPM+** | **-1.721** | **2.629** | **10.306** | **0.962** | **2.433** |
| BAPM | -1.667 | 2.815 | 11.035 | 0.958 | 2.626 |
| WBPM | -1.594 | 2.978 | 11.672 | 0.937 | 3.003 |
| WBPM-OLS | -1.622 | 3.294 | 12.913 | 0.948 | 3.257 |
| MH | -1.634 | 3.349 | 13.127 | 0.946 | 3.275 |
| CR | -1.698 | 3.402 | 13.335 | 0.942 | 3.479 |
| CR+ | -1.707 | 2.814 | 11.029 | 0.941 | 2.829 |

Table 5: Metrics of Simulations Using Data Based on the CNN Study (Averaged over 5,000 iterations), N = 200. BAPM+: Batch-Adaptive Pair Matching with Regression Adjustment; BAPM: Batch-Adaptive Pair Matching; WBPM: Within-Batch Pair Matching; WBPM-OLS: Within-Batch Pair Matching with OLS; MH: Mahalanobis Matching; CR: Complete Randomization; CR+: Complete Randomization with Regression Adjustment.

| Method | ATE | SE | CI Length | Coverage | RMSE |
| --- | --- | --- | --- | --- | --- |
| **BAPM+** | **-1.620** | **1.683** | **6.598** | **0.955** | **1.577** |
| BAPM | -1.587 | 1.808 | 7.089 | 0.952 | 1.746 |
| WBPM | -1.570 | 2.036 | 7.980 | 0.969 | 1.847 |
| WBPM-OLS | -1.575 | 2.279 | 8.933 | 0.943 | 2.234 |
| MH | -1.510 | 2.279 | 8.933 | 0.965 | 2.172 |
| CR | -1.497 | 2.374 | 9.305 | 0.959 | 2.287 |
| CR+ | -1.505 | 1.840 | 7.214 | 0.965 | 1.722 |